\documentclass[aps,prl,twocolumn,amsmath,amssymb,superscriptaddress,letterpaper]{revtex4-1}

\usepackage{times}
\usepackage{amsfonts}
\usepackage{mathrsfs}
\usepackage{graphicx}
\usepackage{dcolumn}
\usepackage{bm}
\usepackage{color}

\usepackage[colorlinks,bookmarks=false,citecolor=blue,linkcolor=red,urlcolor=blue]{hyperref}

\def\be{\begin{equation}}       \def\ee{\end{equation}}
\def\bea{\begin{eqnarray}}      \def\eea{\end{eqnarray}}

\begin{document}
\title{Topological magnetic line defects in Fe(Te,Se) high-temperature superconductors }

\author{Xianxin Wu}
\email{xianxinwu@gmail.com}
\affiliation{Department of Physics, the Pennsylvania State University, University Park, PA, 16802}
\affiliation{Beijing National Laboratory for Condensed Matter Physics,
and Institute of Physics, Chinese Academy of Sciences, Beijing 100190, China}

\author{Jia-Xin Yin}
\affiliation{Beijing National Laboratory for Condensed Matter Physics,
and Institute of Physics, Chinese Academy of Sciences, Beijing 100190, China}
\affiliation{Department of Physics, Princeton University, Princeton, New Jersey 08544,  USA}

\author{Chao-Xing Liu}
\email{cxl56@psu.edu}
\affiliation{Department of Physics, the Pennsylvania State University, University Park, PA, 16802}

\author{Jiangping Hu}
\email{jphu@iphy.ac.cn}
\affiliation{Beijing National Laboratory for Condensed Matter Physics,
and Institute of Physics, Chinese Academy of Sciences, Beijing 100190, China}
\affiliation{CAS Center of Excellence in Topological Quantum Computation and Kavli Institute of Theoretical Sciences,
	University of Chinese Academy of Sciences, Beijing 100190, China}

\date{\today}

\begin{abstract}
 Magnetic impurity chains on top of conventional  superconductors are promising platforms to realize Majorana modes. Iron-based high-temperature superconductors are known in the vicinity of magnetic states due to the strong Hund's coupling in iron atoms. Here we propose that the line defects with missing Te/Se anions in Fe(Se,Te) superconductors provide the realization of intrinsic antiferromagnetic(AFM) chains with Rashba spin-orbit coupling.  Against conventional wisdom, Majorana zero modes (MZMs) can be robustly generated at these AFM chain ends. These results can consistently explain the recent experimental observation of zero-energy end states in line defects of monolayer Fe(Te,Se)/SrTiO$_3$ by scanning tunneling microscopy (STM) measurements. Our research not only demonstrates an unprecedented interplay among native line defect, emergent magnetism and topological superconductivity but also explores a high-temperature platform for Majorana fermions.

\end{abstract}

\maketitle

\textit{Introduction}   Majorana zero modes, hosted in the surface or edge of topological superconductors, have drawn enormous attentions in condensed matter physics, due to its non-Abelian statistics, which is essential for  topological quantum computation\cite{Ivanov2001,Kitaev2003,Kitaev2006,Nayak2008,Alicea2012,Sarma2015,Aasen2016,Elliott2015RMP,Karzig2017}. There have been many studies including both theoretical proposals \cite{RiceTM1995,DasSarma2006,Fu2008,Lutchyn2010,Oreg2010,Sau2010,AliceaJ2010,Nadj-Perge2013PRB,Braunecker2013} and experimental efforts \cite{Mourik2012,WangMX2012,Nadj-Perge2014,Sun2016,Deng2016,ZhangH2018,Lutchyn2018} for their realization. In particular, a ferromagnetic atomic chain on an $s$-wave superconducting substrate\cite{Nadj-Perge2014} has been experimentally shown to generate MZMs at its ends, where the spin-polarized bands are forced to favor $p$-wave pairing. However, there is little investigation along the other way of thinking, namely, searching for superconductors with intrinsic magnetic chains. As conventional superconductors are incompatible with magnetism, unconventional superconductors are promising candidates.

Recently, theoretical predictions and experimental measurements have identified topological band structures in some families of iron-based superconductors\cite{Hao2014,Wu2016,Wang2015,XuPRL2016,Zhang2018,Zhang2019NP,HaoNSR2019,Shi2017,Peng2019}. The natural integration of topological properties and high T$_c$ superconductivity in iron based superconductors have rendered them an exciting platform to realize topological superconductivity at high temperature. MZMs localized in impurities or vortex cores are evidenced by the zero-bias peaks in STM experiments in both iron chalcogenides (Fe(Te,Se) crystals) and iron pnictides (CaKFe$_4$As$_4$)\cite{YinJX2015,WangDF2018,LiuQ2018,KongLY2019,Machida2019,LiuWY2019,ZhangST2020}. In addition, in two-dimensional
(2D) monolayer FeSe$_{1-x}$Te$_x$/SrTiO$_3$ (STO), the band inversion process at $\Gamma$ point has been directly observed with increasing $x$ and the system becomes topologically nontrivial when $x>0.79$\cite{Shi2017,Peng2019}. Based on discovered topological band structures, high-order topological superconductivity with Majorana hinge/corner states has been proposed to be realized in iron based superconductors as well \cite{WangQY2018,YanZB2018,ZhangRX2019PRL,Wufetese2019,ZhangRX-PRL20192}.

Besides the topological properties, one of the most prominent features for iron-based superconductors, distinct from conventional superconductors, is that they are in a vicinity of magnetic order states owing to the strong Hund's coupling in iron atoms. Despite an isolated Fe atom has a large magnetic moment, in the crystals of Fe-based superconductors, electrons of Fe atoms become delocalized through hybridizations with anions, suppressing local magnetic moments. Thus, in the absence of or by weakening the anion bridging, Fe atoms have a tendency towards strong local magnetism. Line defects formed by missing anions have been observed recently in a monolayer FeTe$_{0.5}$Se$_{0.5}$/STO\cite{ChenC2020}. Surprisingly, a zero-bias peak at the ends of atomic line defects was detected, highly resembling the characteristics of MZMs\cite{ChenC2020}. Considering the magnetic nature of Fe atoms, it is naturally to conjecture  that the line defects  may be new  platforms for high-temperature MZMs.

 In this work, we study the electronic properties of these line defects formed by missing Se/Te anions in Fe(Se,Te) monolayer to explore their topological nature. Our first-principles calculations reveal that the $d_{yz}$ and $d_{x^2-y^2}$ orbitals of Fe atoms in the line defect contribute to flat bands near the Fermi level, leading to a magnetic instability. Further calculations suggest an AFM configuration is energetically more favorable, in sharp contrast to the hypothetical nonmagnetic nature. In both ferromagnetic(FM) and AFM configurations, $d_{xz}$ bands are partially occupied and dominantly contribute to Fermi surfaces.  By including Rashba spin-orbit coupling (SOC), an odd number of 1D bands cross the Fermi level  in the magnetic states and the underlying intrinsic superconductivity in Fe(Se,Te) drives the line defect into a 1D topological superconducting phase with MZMs at its ends. To our knowledge, this is  the first realistic instance of realizing MZMs in an AFM chain. Owing to the compatibility of superconductivity and antiferromagnetism, our study suggests that the missing anion magnetic line defects provide an unique platform to explore AFM topological superconductivity and MZMs.

\begin{figure}[tb]
\centerline{\includegraphics[width=1.0\columnwidth]{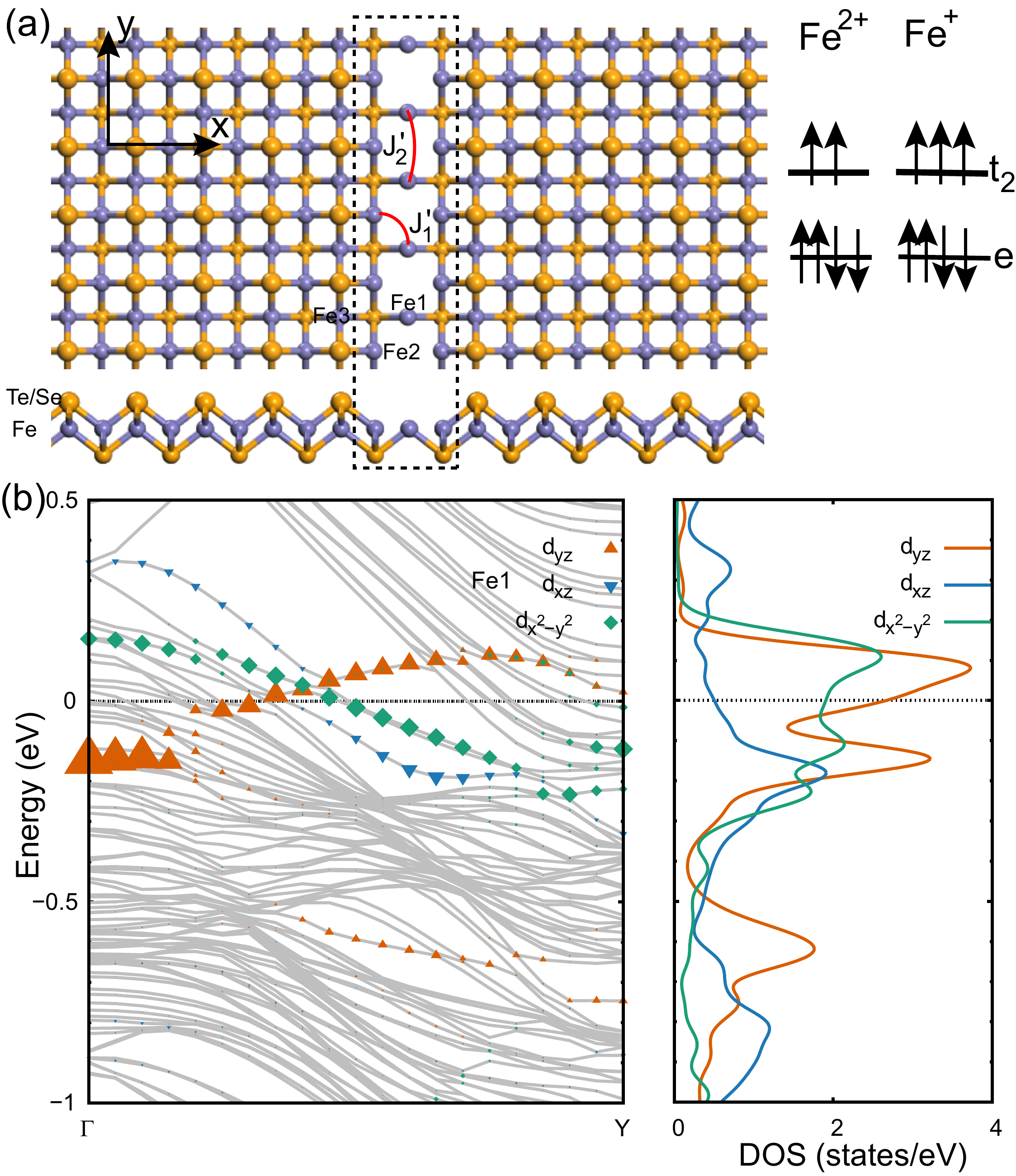}}
\caption{(color online) (a) A schematic of the line defects in monolayer Fe(Te,Se)/STO and occupations of Fe$^{2+}$ and Fe$^+$ $3d$ electrons in tetrahedral coordination environments. (b) Band structure and DOS for the line defects in monolayer FeSe. The sizes of triangles and diamonds represent the orbital weight of iron atoms Fe1 in the line defect.
 \label{crystal} }
\end{figure}

\begin{figure}[tb]
\centerline{\includegraphics[width=1.0\columnwidth]{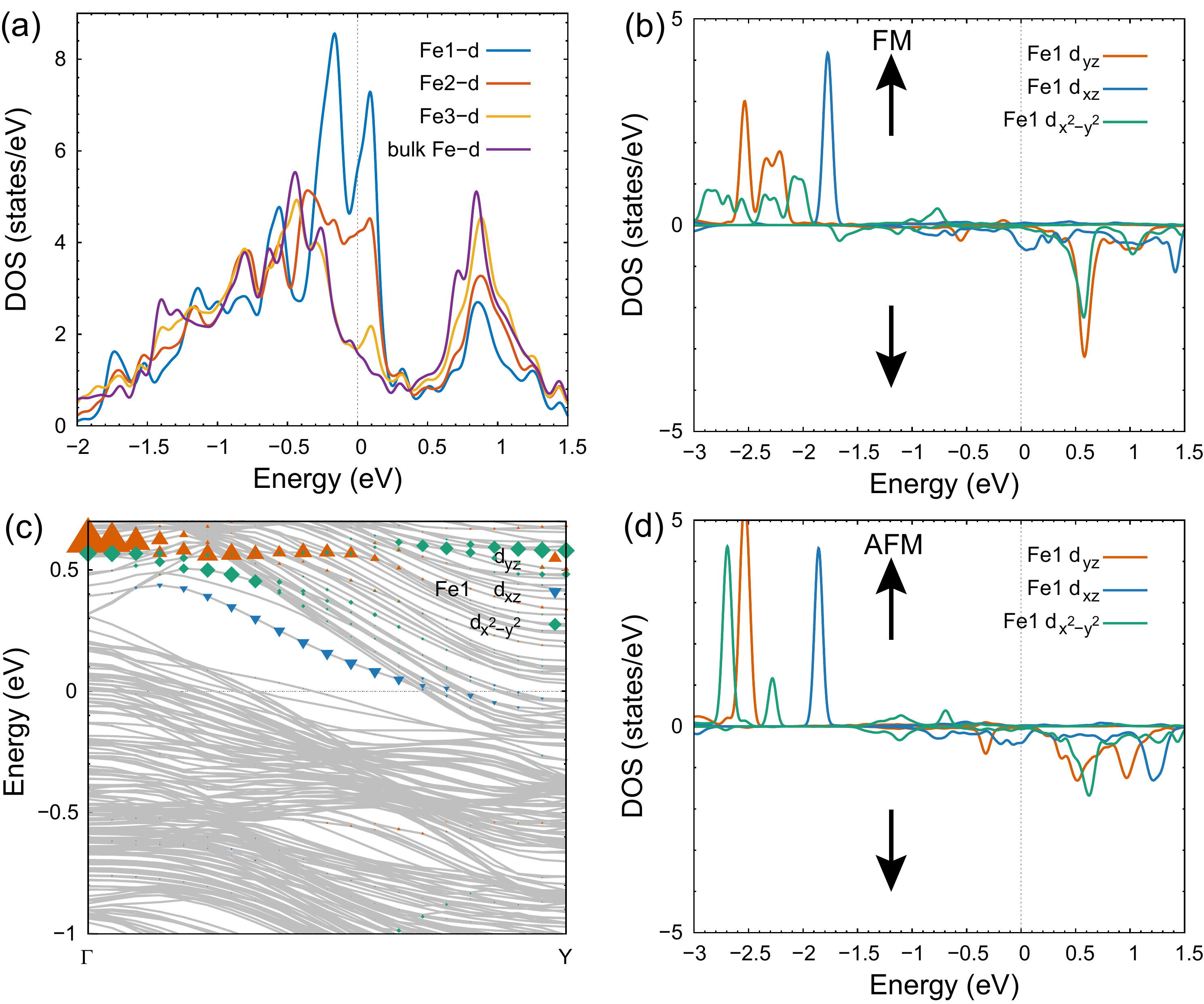}}
\caption{(color online) Density of states for Fe atoms and band structures for line defects in monolayer Fe(Te,Se) in nonmagnetic, FM and AFM states. (a) DOS for Fe atoms close to the line defect in a nonmagnetic state. (b) Spin-resolved DOS and (c) orbital-resolved band structure for Fe1 atoms in a FM state. (d) Spin-resolved DOS for one Fe1 atom in an AFM state. The sizes of triangles and diamonds represent the  minority-spin orbital weight of Fe1 atom in the line defect.
 \label{banddos} }
\end{figure}

\textit{Band structure for line defects in monolayer Fe(Te,Se)} The line defect in monolayer Fe(Te,Se), displayed in Fig.\ref{crystal}(a), corresponds to a line missing of top Te/Se atoms, naturally emerging in the growth process\cite{ChenC2020}. Compared with normal Fe atoms in Fe(Te,Se), the iron atoms Fe1 in the defect can only couple with two nearest Te/Se atoms, which should generate a significant change in the local electronic structure.

We perform first-principles calculations to study the electronic structure for the line defect. In the calculations, the inplane lattice constant $a=3.905$\AA~ and the height of the Se/Te anions from the Fe plane $h=1.50$\AA~ are adopted\cite{Peng2019}. To model a line defect in Fe(Te,Se), we choose a $15\times 1$ supercell of monolayer FeSe with a vacuum layer of 25 \AA~ along $z$ direction and remove a line of top Se atoms in the center. This slab is large enough to avoid interactions between the adjacent line defects.

In the following, we discuss the line defect in monolayer FeSe in our calculations, as the atomic relaxation and the substitution of Te for Se will not qualitatively change the results (see supplementary materials (SM)). Due to the tetrahedral crystal field in iron based supercondutors, the five $d$ orbitals of iron are split into $t_2$ and $e$ orbitals. In Fe(Te,Se) systems, iron atoms have a nominal valence of Fe$^{2+}$ and $e$ orbitals are occupied while $t_2$ orbital are partially filled, contributing dominantly to the Fermi surfaces. However, the absence of top Te/Se atoms in the line defect will change the valence of the corresponding iron atoms Fe1, which are expected to have a nominal 3d$^7$ (Fe$^+$) configuration.  Fig.\ref{crystal} (b) displays the band structure and density of states for a line defect in monolayer FeSe. We notice that the $t_2$ orbitals of Fe1, including $d_{xz}$, $d_{yz}$ and $d_{x^2-y^2}$, are nearly half-filled, consistent with occupations of Fe$^+$ $3d$ electrons. Moreover, the most prominent feature is that the $d_{yz}$ and $d_{x^2-y^2}$ bands of Fe1 are extremely flat, in sharp contrast to the normal Fe bands. These flat bands originate from the reduced hopping along $y$ direction due to the missing line Te/Se atoms. While the $d_{xz}$ band has a larger bandwidth as the indirect hopping can still survive through the nearest Te/Se and Fe atoms. These flat bands contribute a large density of states (DOS) near the Fermi level, as displayed in Fig.\ref{crystal} (b). Compared with electron pockets around M point and hole bands around $\Gamma$ for monolayer Fe(Te,Se)/STO in ARPES experiments\cite{Shi2017}, the Fermi level crossings of hole bands around $\Gamma$ point in our calculations are ascribed to the neglect of electron doping from the substrates and inadequate treatment of correlation effect in first-principles calculations.

We further display the DOS of Fe atoms close to the line defect in Fig.\ref{banddos}, where the labels of Fe atoms can be found in Fig.\ref{crystal}(a). The DOS at the Fermi level $D(E_f)$ of Fe atoms in the line defect is almost two times larger than that of bulk Fe. This large $D(E_f)$ is relatively robust against electron doping in the realistic monolayer Fe(Te,Se)/STO. According to the Stoner criterion, a large $D(E_f)$ can induce a magnetic instability if interactions are sufficiently strong. The strong interactions from Fe 3d nature suggest a magnetic line defect in monolayer Fe(Te,Se). Despite possible spin fluctuations in monolayer Fe(Te,Se), so far, there are no evidences demonstrating any static magnetic order in monolayer Fe(Te,Se).

\textit{Magnetic configurations of line defects} We turn to investigate the magnetic order and the corresponding electronic structures of line defects in monolayer Fe(Te,Se). The magnetic order in iron chalcogenides can be described by a Heisenberg model with the nearest, the next-nearest, and the next next-nearest neighbor couplings $J_1$, $J_2$, and $J_3$\cite{Ma2009}. All of them have a superexchange origin mediated by Te/Se hence are antiferromagnetic. As a consequence, the absence of top Te/Se atoms in the line defect will significantly reduce the corresponding exchange coupling between nearest and next-nearest neighbor Fe1 atoms and we label these exchange couplings as $J'_1$ and $J'_2$, as depicted in Fig.\ref{crystal}(a). $J'_1$ is approximately reduced to half of the original value, i.e., $J'_1=J_1/2$.  For Fe1 with half-filled $t_2$ orbitals, $J'_2$ is derived from the direct exchange coupling and should be antiferromagnetic.

For this one-dimensional (1D) line defect, we only consider FM and AFM configurations and neglect complicated spiral magnetic orders due to the short-ranged exchange couplings. In the FM configuration, the antiferromagnetic $J'_1$ can induce a small opposite magnetic moment on Fe2 with respect to Fe1 (see SM). The magnetic states of the defect can be described by a Heisenberg model with nearest neighbor coupling $J_{eff}$,
\begin{eqnarray}
H=\sum_{\langle ij \rangle} J_{eff}\bm{S}_{1i}\cdot\bm{S}_{1j},
\end{eqnarray}
where $\bm{S}_{1i}$ is the magnetic moment for Fe1 and $J_{eff}$ includes contributions from the direct coupling $J'_2$ and indirect coupling $J'_1$. As $J'_1$ effectively contributes a ferromagnetic coupling between nearest neighbor iron atoms, the $J'_1$ coupling will compete with $J'_2$ term.
 The corresponding energies per Fe for FM and AFM states based on the above Heisenberg model are: $E_{FM}=-4J'_1 S_1S_2+J'_2 S^2_1$, $E_{AFM}=-J'_2 S^2_1$, where $S_{1}$ ($S_2$) is the magnetic moment for iron atoms Fe1 (Fe2) in the line defect. The line defect favors a FM order if $J'_2/J'_1<2S_2/S_1$ otherwise an AFM order. According to our calculation, the magnetic states have a much lower energy compared with the nonmagnetic state (about 1 eV/Fe) and $S_2/S_1\sim 0.1$ in the FM state. Moreover, the AFM configuration is 13 meV/Fe lower in energy than the FM configuration, leading to $J_{eff}=6.5 meV/S^2_1$, where the magnetic moment of Fe1 is about 2.9 $\mu_B$, close to the value for half-filled $t_2$ orbitals.  The easy axis of the
Fe spin moments is out-of-plane ($z$ axis in Fig. \ref{crystal}) and
about 1.7 meV/Fe lower in energy than the two high-symmetry in-plane directions.

 In the FM state, the DOS for Fe1 atoms and band structure is shown in Fig.\ref{banddos}(c), where majority-spin $t_2$ orbital are occupied while the minority-spin $d_{xz}$ band is partially filled, contributing two Fermi points around Y. In the AFM state, each band is two-fold degenerate without SOC, the minority-spin $d_{xz}$ band of one Fe1 is partially occupied, as shown in Fig.\ref{banddos}(d),  and there are Fermi points around Y (see SM).

\begin{figure*}[tb]
\centerline{\includegraphics[width=2.0\columnwidth]{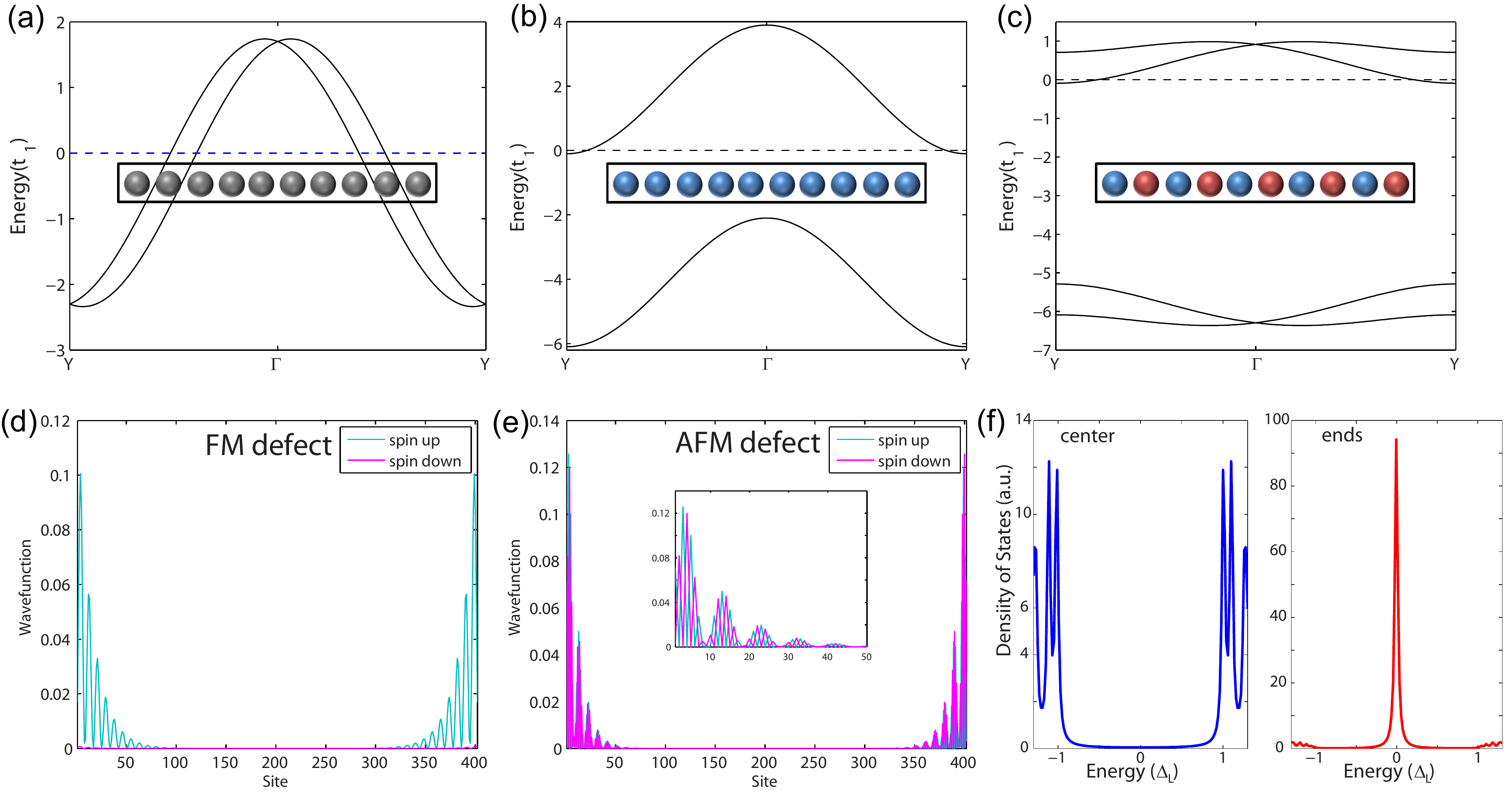}}
\caption{(color online) Band structures in normal states for the line defect with nonmagnetic (a), FM (b) and AFM (c) states. The spin-resolved wavefunctions of MZMs in the FM (d) and AFM (e) line defects. (f) Local DOS at the center and ends of the AFM line defect including superconductivity with open boundary condition. The electron filling is $n=1.1$ per site and we adopt $\lambda_R=0.2 t_1$ and $\bm{m}=6 t_1\hat{\bm{z}}$ ($\bm{m}_1=6t_1\hat{\bm{z}}$) for the FM (AFM) line defect. $\Delta_L$ is the effective pairing gap for the band crossing the Fermi level in AFM line defects.
 \label{MZMdos} }
\end{figure*}

\textit{Majorana modes at the ends of line defects} From our above calculations, we find that the line defects in monolayer Fe(Te,Se) are intrinsically magnetic. Although the energy of AFM configuration is slightly lower than that of FM configuration, we consider both cases in the following model calculations to investigate the topological superconductivity in them. The 1D line defect can be theoretically described by the Hamiltonian,
\begin{eqnarray}
\mathcal{H}_{LD}&=&\sum_{ ij\alpha\sigma}[t_{\alpha,ij}\delta_{\langle ij \rangle}+(\epsilon_\alpha-\mu) \delta_{ij}]c^{\dag}_{i\alpha\sigma}c_{j\alpha\sigma}\nonumber\\
&+&\sum_{ i\alpha\sigma_1\sigma_2}\frac{1}{2}\bm{M}_i\cdot\bm{\sigma}_{\sigma_1\sigma_2}c^{\dag}_{i\alpha\sigma_1}c_{i\alpha\sigma_2}\nonumber\\
&+&\sum_{ \langle ij\rangle\alpha}[i\lambda_R(c^{\dag}_{i\alpha\uparrow}c_{j\alpha\downarrow}+c^{\dag}_{i\alpha\downarrow}c_{j\alpha\uparrow})+h.c.]\nonumber\\
&+&\sum_{ij\alpha\sigma}[\sigma(\Delta_0\delta_{ij}+\Delta_1\delta_{\langle ij \rangle})c^{\dag}_{i\alpha\sigma}c^\dag_{j\alpha\bar{\sigma}}+h.c.],
\end{eqnarray}
where $\langle ij \rangle$ labels the  nearest neighbor Fe1 sites, $\alpha=1,2,3$ represent $d_{xz}$, $d_{yz}$ and $d_{x^2-y^2}$ orbitals for iron atoms in the line defect and $\sigma_i$ labels the Pauli matrix in spin space. The first term is the kinetic energy part and the second term describes the magnetic coupling for each iron site, originating from the magnetization of $d_{yz}$ and $d_{x^2-y^2}$ orbitals. This coupling can be expressed as $\bm{M}_i=\bm{m}_0+(-1)^i\bm{m}_1$, where $\bm{m}_0$ ($\bm{m}_1$) is the FM (AFM) coupling. The third term is the Rashba SOC due to the mirror symmetry breaking from the absence of top Se/Te atoms, which is sizable in first-principles calculations (see SM) and crucial for topological superconductivity in the AFM configuration. The last term describes the onsite and nearest neighbor spin singlet pairing in proximity to the superconducting monolayer Fe(Te,Se) and the orbital independent pairing is adopted for convenience. By fitting to the band structure in Fig.\ref{crystal}(b), the hopping parameters are $t_{1}=0.3$, $t_2=-0.13$ and $t_3=0.18$ eV.

The band structures from the above model in magnetic states are qualitatively consistent with first-principle calculations in both configurations (see SM). Since the Fermi points in normal states are predominantly attributed to $d_{xz}$ orbitals, we can further simplify the above model to a single-orbital one. Without magnetic orders, the representative band is shown in Fig.\ref{MZMdos}(a), where time reversal symmetry protects a degeneracy at $k=0,\pi$ and the Fermi level always crosses even number of bands. In the FM state with time reversal symmetry breaking, the bands are non-degenerate and only one minority-spin band with crosses the Fermi level, contributing two Fermi points around Y, as shown in Fig.\ref{MZMdos} (b). As the electron pockets of monolayer Fe(Te,Se) are projected to Y point in line defects, these Fermi points will obtain a superconducting gap by proximity effect. Most importantly, the effective pairing is $p$-wave  as required by the Fermionic antisymmetry, resembling to the 1D Kitaev model. Due to the nontrivial topology, the system remains gapped in the center while two MZMs will occur at the ends of the line defect (see SM).

In the AFM state, there is an effective time reversal symmetry $\tilde{\Theta}=\Theta T_{\frac{1}{2}}$, combining time reversal symmetry $ \Theta$ and a half-lattice translation $T_{\frac{1}{2}}$. $\tilde{\Theta}^2=\mp 1$ at $k=0,\pi$ suggests that Kramers degeneracy only occurs at $k=0$ but not $k=\pi$. At a generic $k$ point including $k=\pi$, the intrinsic Rashba SOC lifts the degeneracy. In contrast to the Rashba band of a nanowire, the prominent feature is that an odd number of 1D bands cross the Fermi level in a wide range of chemical potential. When the exchange coupling $m_1$ is relatively strong (from our first-principles calculations), the representative band structure is displayed in Fig.\ref{MZMdos}(c) and only one band with mixed minority-spin and majority-spin contributions cross the Fermi level. Further including superconductivity, it naturally induces MZMs located at the ends of the line defect, displayed in Fig.\ref{MZMdos} (f). We emphasize that there is only one MZM at each end, distinct from the topological Shockley defect scenario\cite{ChenC2020}. Fig.\ref{MZMdos}(d) and (e) display the spin-resolved spatial profiles of MZMs in FM and AFM line defects. The MZMs are localized at ends in both cases but the spin polarization of MZMs differs. In the former, there is a uniform spin polarization, however, in the latter, the spin polarization is spatially alternating.

\begin{figure}[tb]
\centerline{\includegraphics[width=1.0\columnwidth]{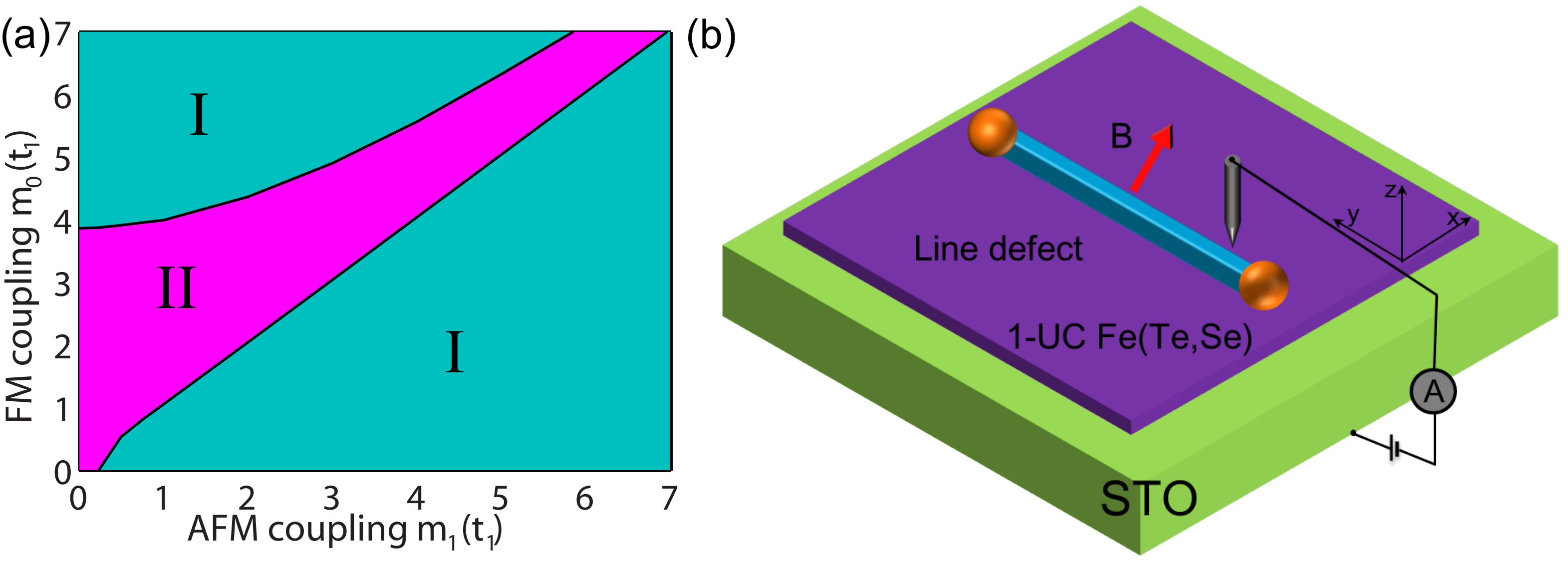}}
\caption{(color online) (a) Phase diagram for topological superconductivity of line defect as a function of AFM and FM couplings. In region I (II), an odd (even) number of 1D bands cross the Fermi level, resulting a topologically nontrival (trivial) phase. The electron filling is $n=1.1$ per site and we adopt $\lambda_R=0.2 t_1$ and the exchange field is perpendicular to the line defect(along $z$ direction). (b) The schematic for experimentally distinguishing topological Shockley defects, AFM and FM scenarios.
 \label{phasediagram} }
\end{figure}

The phase diagram of topological superconductivity in line defect as a function of AFM and FM couplings is displayed in Fig.\ref{phasediagram}(a). In blue regions, an odd number of bands cross the Fermi level, inducing a topological phase. Furthermore, the $m_1$-dominated region is much larger than the $m_0$-dominated one, suggesting that AFM line defect is easier for the realization of MZMs. In the pink region, there are an even number of bands crossing the Fermi level and the system is generally topologically trivial in the class D, where the topological invariant is $\mathbb{Z}_2$ in 1D. The $m_0$-dominated and $m_1$-dominated phases are always separated by a trivial from a band analysis (see SM). If an increasing external magnetic field can induce a phase transition from an AFM state to a FM state, it could also render topological phase transitions first from a nontrivial phase to a trivial one then back to a nontrivial one. The corresponding local DOS evolution at ends of line defects can be found in SM.

\textit{Discussion} Our results are distinct from those in Ref.\cite{ChenC2020}, where MZMs are considered as a Kramers pair in topological Shockley defects protected by time-reversal symmetry. These two different schemes can be distinguished by applying an external magnetic field. In the Shockley case, once the time-reversal symmetry is broken by an external magnetic field parallel to the Rashba spin-orbit field (along $x$ axis in Fig.\ref{phasediagram}(b)), the Krammers pair of MZM will split. In our case, however, there is only one MZM at each end of the magnetic line defect and it is robust against weak external perturbations.

The scenario of the AFM line defects is consistent with available experimental evidences. The observed in-gap bound states on a single Te/Se vacancy defect in monolayer Fe(Te,Se)/STO are reminiscent of the Yu-Shiba-Rusinov (YSR) states in superconductors\cite{ChenC2020}, indicating its magnetic nature. If the line defect is FM,  YSR states are also expected. However, there are no other in-gap states except zero-energy end states on the line defect in STM measurements\cite{ChenC2020}. In contrast, as the total magnetization of an AFM line defect vanishes, it performs as a nonmagnetic impurity and therefore there are no in-gap states, which is consistent with STM measurements. Moreover, the effective superconducting gap for the line defect is proportional to the Rashba SOC strength in the FM state and is expected to be much smaller than the bulk superconducting gap. On the contrary, the nearest-neighbor pairing is allowed in the AFM state and the SC gap should be comparable to the bulk value, which is consistent with experimental measurements\cite{ChenC2020}.

To experimentally distinguish AFM and FM scenarios shown in Fig.\ref{phasediagram} (b), spin-polarized STM measure can provide direct evidence about the magnetic nature of line defects and the spin-resolved spatial profiles of MZMs at ends, as depicted in Fig.\ref{MZMdos} (d) and (e). Moreover, MZMs in  the two cases will exhibit distinct behaviors under an external magnetic field $\bm{B}$ along the magnetization axis: the MZMs are robust in FM line defects; however, for AFM line defects, a large magnetic field drives topological phase transitions and the zero-bias peak at ends will first split at certain $\bm{B}$ and emerge again at a larger $\bm{B}$ (see SM).

\textit{Conclusion} We study the topological superconductivity of intrinsic line defects in monolayer Fe(Te,Se)/SrTiO$_3$. First-principles calculations reveal that the missing Te/Se atoms introduce a large DOS near the Fermi level, inducing a magnetic order on the line defect. In either FM or AFM configurations, the line defect is 1D topologically superconducting, which hosts MZMs at its ends, consistent with recent STM experiments. In particular, we find the AFM configuration is energetically more favorable and the MZMs in the AFM configuration has a spatially alternating spin-polarized profile. The AFM line defects, derived from atomic vacancies, are quite common and almost unavoidable and can also occur in other iron based superconductors or superconducting materials. As superconductivity and antiferromagnetism are compatible, they offer a novel and concise platform to explore AFM topological superconductivity and realize MZMs.

{\it Acknowledgement: }
C.X.Liu acknowledges the support of the Office of Naval Research (Grant No. N00014-18-1-2793), DOE grant (DE-SC0019064) and Kaufman New Initiative research grant of the Pittsburgh Foundation.
J.P.Hu was supported by the Ministry of Science and Technology of China 973 program
(No. 2017YFA0303100), National
Science Foundation of China (Grant No. NSFC11888101), and the
Strategic Priority Research Program of CAS (Grant
No.XDB28000000).

\bibliography{linedefectref}

\end{document}